\documentclass[sigconf]{acmart}

\renewcommand\footnotetextcopyrightpermission[1]{} 
\AtBeginDocument{%
  }

\renewcommand\footnotetextcopyrightpermission[1]{} 
\begin{document}

\title{Convolutional neural network classification of cancer cytopathology images: taking breast cancer as an example\\ 
	}

\author{MingXuan Xiao}
\affiliation{%
  \institution{SouthWest JiaoTong University}
  \country{China}
  }
\email{553556963albert@gmail.com}

\author{Yufeng Li}
\affiliation{%
  \institution{University of Southampton}
  \country{China}
  }
\email{liyufeng0913@gmail.com}

\author{Xu Yan}
\affiliation{%
  \institution{Trine University}
  \country{USA}
  }
\email{xyan232@my.trine.edu}

\author{Min Gao }
\affiliation{%
 \institution{Trine University}
 \country{USA}
  }
\email{mingao4460@gmail.com}

\author{Weimin Wang}
\affiliation{%
  \institution{Hong Kong University of Science and Technology }
  \country{China}
  }
\email{wangwaynemin@gmail.com}

\renewcommand{\shortauthors}{}

\begin{abstract}
Breast cancer is a relatively common cancer among gynecological cancers. Its diagnosis often relies on the pathology of cells in the lesion. The pathological diagnosis of breast cancer not only requires professionals and time, but also sometimes involves subjective judgment. To address the challenges of dependence on pathologists expertise and the time-consuming nature of achieving accurate breast pathological image classification, this paper introduces an approach utilizing convolutional neural networks (CNNs) for the rapid categorization of pathological images, aiming to enhance the efficiency of breast pathological image detection. And the approach enables the rapid and automatic classification of pathological images into benign and malignant groups. The methodology involves utilizing a convolutional neural network (CNN) model leveraging the Inceptionv3 architecture and transfer learning algorithm for extracting features from pathological images. Utilizing a neural network with fully connected layers and employing the SoftMax function for image classification. Additionally, the concept of image partitioning is introduced to handle high-resolution images. To achieve the ultimate classification outcome, the classification probabilities of each image block are aggregated using three algorithms: summation, product, and maximum. Experimental validation was conducted on the BreaKHis public dataset, resulting in accuracy rates surpassing 0.92 across all four magnification coefficients (40X, 100X, 200X, and 400X). It demonstrates that the proposed method effectively enhances the accuracy in classifying pathological images of breast cancer.
\end{abstract}



\maketitle
    \section{Introduction}
Among gynecological cancer diseases, breast cancer stands as one among the predominant malignant tumors affecting women's health and holds the leading position in female malignant tumor-related mortalities. Its incidence exhibits significant regional disparities globally, with markedly higher rates observed in developed countries and regions compared to their less developed counterparts. In 2020, the World Health Organization reported a global incidence of 2.26 million fresh instances of breast cancer, with a high mortality rate of 685,000, both ranking highest in terms of incidence and mortality rates among women worldwide\cite{chhikara2023global}. The latest projections from the American Cancer Society indicate that by 2023, the count of recently diagnosed cases of female breast cancer in the United States is anticipated to approach nearly 300,000. Histopathological analysis remains the predominant methodology for diagnosing breast cancer, with the majority of diagnoses reliant on visual examination of histological specimens under a microscope, necessitating substantial expertise from specialized pathologists\cite{cserni2020histological}. The diagnostic concordance among experts averages approximately 75\%, underscoring the considerable dependence on the experiential acumen of pathologists for accurate breast cancer detection\cite{elmore2015diagnostic}. Consequently, The utilization of computational techniques for the automated categorization of histopathological images not only expedites the diagnostic process for breast cancer but also mitigates the propensity for errors\cite{liu2023unveiling}\cite{10440308}. The efficient and accurate method is crucial for achieving this objective.

    \section{RELATED WORK}
Currently, research on breast cancer recognition can be broadly categorized into two primary approaches\cite{ma2023implementation}\cite{guo2021multi}: 
1)Methods based on manual feature extraction combined with traditional machine learning. For instance, Belsare employed statistical texture features to train K-NN (k-nearest neighbors) and Support Vector Machine (SVM) classifiers, achieving accuracy rates ranging from 70\% to 100\% on a privately enlarged breast histopathological dataset\cite{belsare2015classification}. Spanhol released the BreaKHis dataset of breast cancer pathology images, investigating the classification performance of 24 groups comprising four classifiers, including SVM, and handcrafted texture features, such as adjacency threshold statistics\cite{spanhol2015dataset}\cite{spanhol2016breast}. As a reference baseline for distinguishing between benign and malignant tumors, the achieved accuracy ranged from 80\% to 85\%. 
2)Classification methods based on deep learning have enabled models to directly extract features from input images, eliminating the demand for manual feature extraction processes and saving considerable human and computational resources\cite{lu2023transflow}. In the past few years, convolutional neural networks (CNNs)\cite{li2018graph}, the key technology of deep learning, has achieved great success in the field of intelligent image recognition and has developed rapidly in the realm of medical examination image analysis\cite{lu2018nuclear}. Zhang \cite{zhang2024enhanced}employed a CNN for the classification of breast cancer pathological images into malignant and non-malignant classes, and successfully identified other tissues, such as normal tissue, benign lesions, carcinoma in situ, and invasive cancer, with an overall accuracy of Also as high as 77.8\%. Spanhol\cite{spanhol2017deep}, utilizing CNNs and deep features on the BreaKHis dataset, classified breast cancer tissue pathology images into benign and malignant classes, achieving maximum accuracies of 90\% and 86.3\%, respectively. Bayramoglu introduced a magnification-agnostic approach for classifying breast cancer tissue pathology images, which simultaneously classifies the pathology image as benign or malignant and identifies the magnification factor. Experimental results from the BreaKHis data set show that the accuracy of distinguishing benign and malignant classification is 84.3\%\cite{bayramoglu2016deep}.
The impact of illumination levels on image recognition was investigated, revealing optimal recognition results at 300lx, while illumination below 200lx led to undetected damage and illumination above 500lx resulted in decreased recognition accuracy\cite{konovalenko2022influence}.To enhance the accuracy of breast pathological image classification, in this study, convolutional neural networks (CNNs) were designed to classify breast cancer pathology images. Additionally, the idea of image segmentation is introduced to tackle high-resolution pathological images. The efficacy of this approach is verified using the BreaKHis-Dataset.

    \subsection{Data Source} 
The utilized dataset in this investigation is the BreakHis dataset, which comprises histopathological images stained with H\&E obtained from tissues of 82 patients, resulting in a total of 7,909 images. Each image in the dataset has a pixel resolution of 700×460, with 2,480 images corresponding to benign samples and 5,429 to malignant samples. The dataset contains images taken at four different magnification levels: 40X, 100X, 200X, and 400X. The distribution of samples across various classes and magnification levels is visually depicted in Figure 1, and Table 1 showed the detailed statistics.

        \begin{table}[h]
        \centering
        \begin{tabular}{|p{2cm}|p{1.5cm}|p{1.5cm}|p{1.5cm}|}
        \hline
        \textbf{Magnification Factor} & \textbf{Benign} & \textbf{Malignant} & \textbf{Total} \\
        \hline
        40X & 625 & 1370 & 1995 \\
        \hline
        100X & 644 & 1437 & 2081 \\
        \hline
        200X & 623 & 1390 & 2013 \\
        \hline
        400X & 588 & 1232 & 1820 \\
        \hline
        total & 2480 & 5429 & 7909 \\
        \hline
        \end{tabular}
        \caption{Distribution of images across different categories and magnification levels}
        \end{table}

        \begin{figure}
            \centering
            \includegraphics[width=0.8\linewidth]{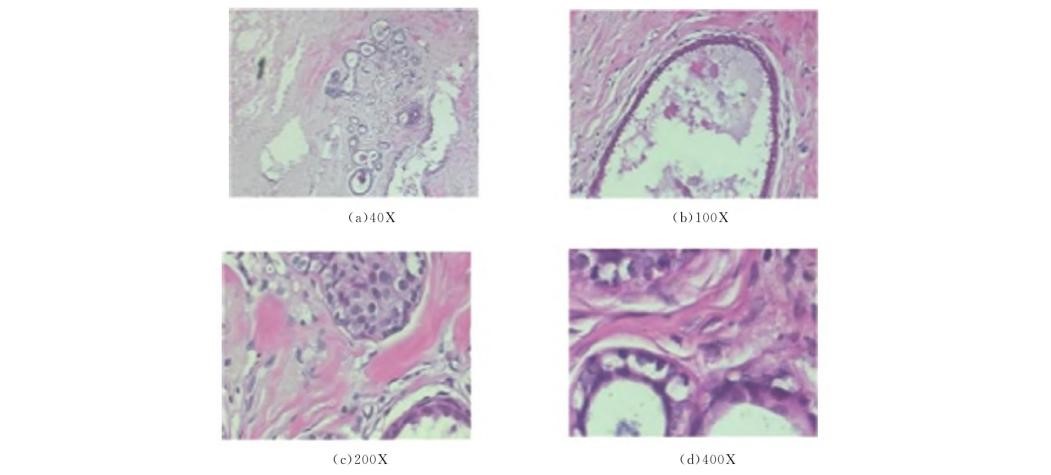}
            \caption{BreakHis dataset samples at various magnification levels}
            \label{fig:enter-label}
        \end{figure}
\section{METHODOLOGY}
Due to the swift progress in deep learning, especially in image recognition, the utilization of deep learning approaches, notably Convolutional Neural Networks (CNNs)\cite{hu2022design}, has outpaced conventional machine learning methods.The CNN architecture is primarily composed of three layers: 
Convolutional layers execute convolution operations on input images employing a series of adjustable filters, producing a feature map for each filter's operation. Fully connected layers, typically positioned at the model's conclusion, amalgamate the extracted features for classification and other tasks.Inception-v3 belongs to the Inception family of CNNs architectures\cite{liu2023ising}. It incorporates various enhancements, such as the implementation of Label Smoothing, the utilization of Factorized 7 x 7 convolutions, and the introduction of an auxiliary classifier. The auxiliary classifier aids in propagating label information to decrease layers of network, complemented by the application of batch normalization for layers in the side head.
This paper employs the Inception-V3 model for the classification and feature extraction of pathological images. Given the model's requirement for images sized at 299×299, the process involves image scaling and feature extraction using the model. The model excludes the last two fully connected layers of the original architecture. The classification process incorporates two new fully connected layers in the neural network, with the first layer comprising 512 nodes and the second layer consisting of 2 nodes. The final layer of the neural network employs the SoftMax function as the activation function for classification.
The specific steps for classifying a single image are as follows:
1)	Resize the image to a fixed size of 299×299. 
2)	Feed the scaled image into the Inception-V3 model for computation. Extract the output parameters of the last pooling layer of the Inception-V3 model as the image's feature vector, with a size of 1×2,048.
3)	Input the feature vector into the fully connected layer neural network for computation. The output results, obtained by applying the SoftMax function to the values of each node in the second layer, indicate the likelihood of the input image being associated with a particular class.
4)	Classify the image into the category with the highest output probability.
This paper adopts the Inception-V3 model parameters trained by ImageNet dataset as initial weights for extracting the feature phase of this model. Fine-tuning approach induces changes in model parameters, requiring the re-calculation of image feature vectors in each training session. This results in a computationally demanding task with prolonged training times. Therefore, the paper employs a fixed weight method. Subsequently, after computing the image feature vector, the paper stores it as a text document. This allows the image's feature vector to be directly retrieved from the text document in the subsequent training process, avoiding redundant computation in the feature extraction process and saving computational time.

\begin{figure}
    \centering
    \includegraphics[width=0.8\linewidth]{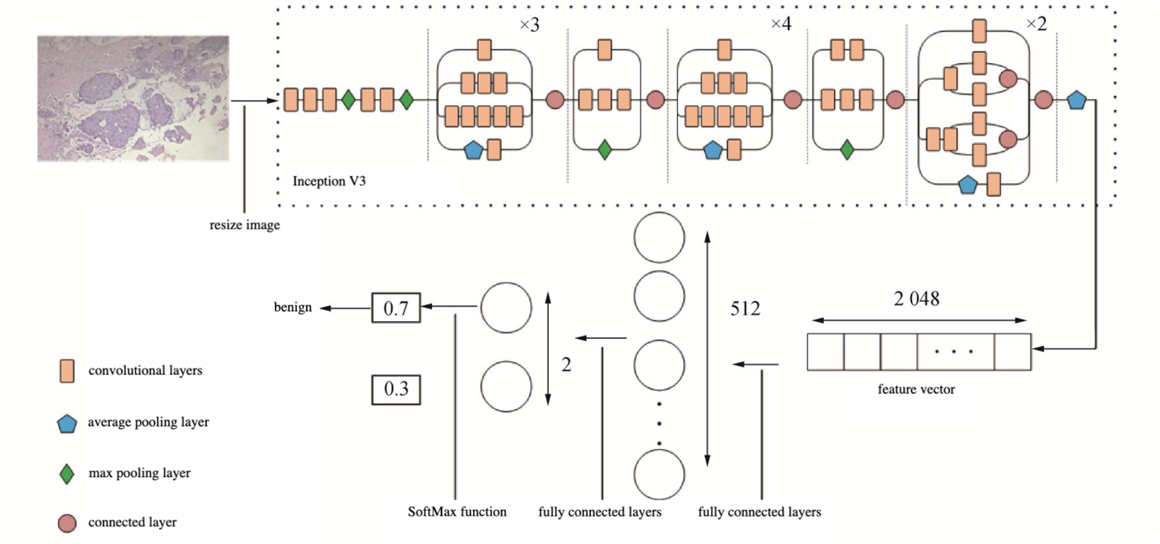}
    \caption{Image Classification based on inception V3 model}
    \label{fig:enter-label}
\end{figure}

The image classification process are illustrated in Figure 2.
During the training process, the cross-entropy between the output results of the SoftMax function and the image labels is constructed as the loss function. Parameter optimization of the fully connected layer neural network is then performed using the gradient descent method to minimize the loss function. The cross-entropy loss function for a single image is defined as (1).

\begin{equation}
    L\left(x,y\right)=-\sum_{k} y_k\log{p_k}\left(x\right)
\end{equation}

Where $p_k(x)$ is the possibility of input image x is classified into class k by the classifier, and $y_k$ is an indicator function such that  $y_k=1$ when the label y of the input image x is class k, otherwise $y_k$=0.
During the training of CNN models, if the dataset has a small number of samples, the network is prone to overfitting. A commonly used method to address this is data augmentation. Data augmentation involves increasing the dataset size through operations such as rotation, flipping, sliding windows, etc. In this research, a data augmentation method established on quadtree segmentation, which is more suitable for high-resolution images, is employed. The quadtree consists of a continuous structure, and at each layer, the input image from the previous layer is evenly segmented into four parts. In the first level, the input image remains the initial image. In the second level, the image is divided into 4 images. In the third level, the image is divided into 16 images, and so on, as illustrated in Figure 3. This paper investigates the results of splitting breast images into the second and third levels, meaning each image is segmented into 4 and 16 pieces, respectively. After segmentation, the training data is increased by 4 times and 16 times the original dataset size for the second and third levels, respectively. After segmentation, each sub-image is considered to have the identical class label as the original image.

\begin{figure}
    \centering
    \includegraphics[width=0.8\linewidth]{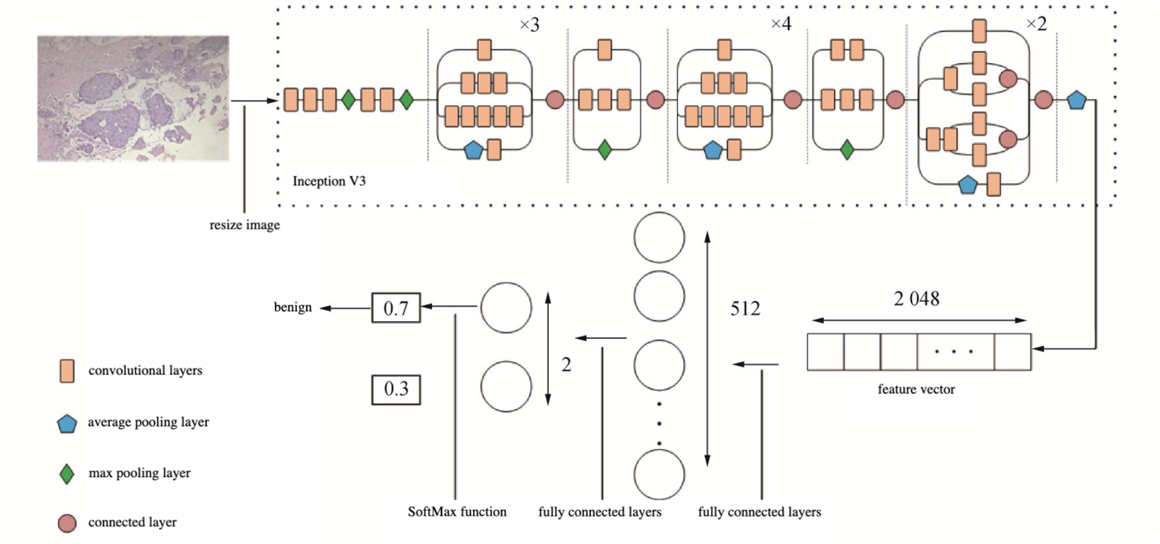}
    \caption{Image segmenting method based on quadtree}
    \label{fig:enter-label}
\end{figure}
When a single image is segmented into multiple sub-image blocks, each block may yield different classification results after model computation. An algorithm is needed to consolidate the classification results of all the sub-image segments. Commonly used algorithms include the sum rule, product rule, maximum rule, majority voting rule, etc. To investigate the impact of different fusion algorithms on breast image classification results, this paper selects and experiments with the sum rule, product rule, and maximum rule. The algorithmic calculation process is shown in (2)(3)(4). Equation (2)  represents the Summation fusion algorithm, equation (3) represents the Product fusion algorithm, and the equation (4) represents the Maximum fusion algorithm.
\begin{equation}
    \phi=\underset{x=1}{\operatorname{argmax}} \sum_{k=1}^{N} p_{k}(x)
\end{equation}
\begin{equation}
\phi=\underset{x=1}{\operatorname{argmax}} \prod_{k=1}^{N} p_{k}(x)
\end{equation}
\begin{equation}
\phi=\underset{x=1}{\operatorname{argmax}} \max _{k=1}^{N} p_{k}(x)
\end{equation}

Where $p_k(x)$ represents the possibility of the ith segmentation being classified into class k by the model. K is the total number of classification categories, and N is the number of blocks into which a single image is divided. Figure 4 illustrates a complete example of the classification process for a pathological image.
\begin{figure}
    \centering
    \includegraphics[width=0.8\linewidth]{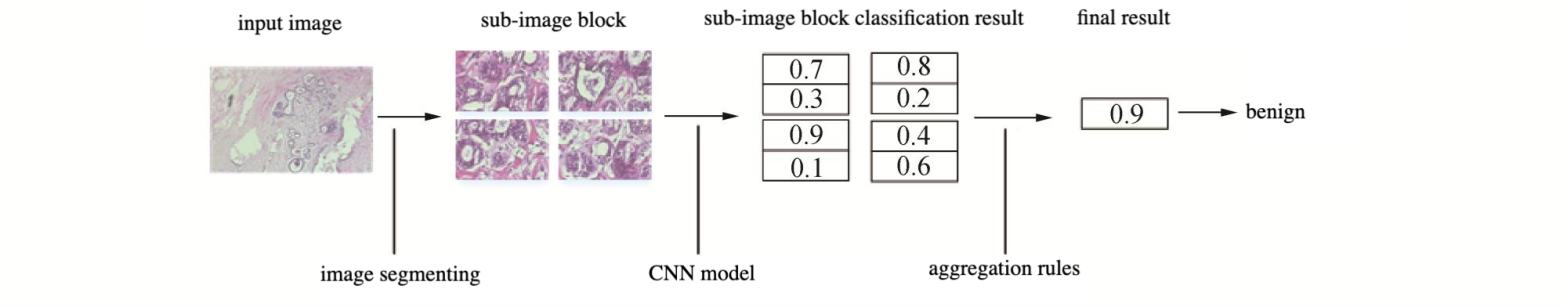}
    \caption{Image classification process}
    \label{fig:enter-label}
\end{figure}

\section{EXPERIMENTAL}
\subsection{Experimental Design}
The dataset is partitioned into three segments: the training set, the validation set, and the testing set. This division is a standard procedure in machine learning and data science. The training dataset used to teach the model to discern specific patterns and relationships in large amounts of data. During the data training phase, the validation set is used to fine-tune the model and adjust relevant hyperparameters. The testing set is used only after the model has been trained, so this dataset must be kept separated from both training set and validation set. In this case, the portion of training set is 75\%, the portion of validation set is 15\%, and the portion of testing set is also 15\%.
The effectiveness of the experiment is measured by classification accuracy, say, the diagnose accuracy. And the diagnose is based on two levels: image-level and patient-level. 
At the image level, every breast pathology image is analyzed and diagnosed into one category.  Thus, the accuracy of image level is simply the division of the number of correctly diagnosed images by the total number of the images as in (5) $N_{correct}$ represents the number of correct diagnoses and  $N_{correct}$ represents the total number of images in the dataset.
\begin{equation}
{Accuracy}_{image}=\frac{N_{correct}}{N_{image}}
\end{equation}
While in the patient level, the diagnosed is based on an individual patient that often comes with several pathology images. So the accuracy of patient level is measured by the average accuracy of each  patient as in (6) $N_{correct}^P$ represents the number of correct diagnoses of patient’s images,  $N_{image}^P$ represents the total number of diagnostic images for patient P and $N_{patient}$ represents the number of patients.
\begin{equation}
{Accuracy}_{patient}=\frac{\sum_{p=1}^{N_{patient}}\frac{N_{correct}^P}{N_{image}^P}}{N_{patient}}
\end{equation}
\subsection{Experimental Results}
Experiments are carried out in two levels, both level of image and patient with the BreaKHis dataset on the same platform. Additional, the result of image level experiment is shown in Table(2) while the result of patient level is shown in Table(3). The accuracy in each table is calculated as 3.1 demonstrated. Both Table(2) and Table(3) shows the segment strategy increases the accuracy of the classification significantly. As for quarter-split, the sample images are divided into four segments, and in the case of 16-way split, the samples are divided into 16 segments. As shown in Table(2), the accuracy of the classification is increased by 2.0\% to 4.4\% on image level compared to the non-split sample. And as shown in Table(3), the accuracy of the classification is increased by 0.9\% to 4.9\% on the patient level compared to the non-split sample. In conclusion, the combination of image segmentation methods with fusion algorithms effectively enhances the accuracy of breast pathological image recognition. It is notable that in 79.2\% cases, the quarter-split segmentation has higher accuracy than the 16-way-split, which is caused by the lack of detail in the process of segmentation.

\begin{table}[h]
\centering
\begin{tabular}{|c|c|c|c|c|c|}
\hline
\multicolumn{1}{|c|}{\textbf{Segmentation}} & \multicolumn{1}{c|}{\textbf{Fusion}} & \multicolumn{4}{c|}{\textbf{Magnification Factor}} \\
\cline{3-6}
& & 40 & 100 & 200 & 400 \\
\hline
Non-split & - & 92.8 & 90.7 & 91.8 & 88.67 \\
\hline
Quarter-split & Summation & 95.0 & 95.1 & 93.8 & 92.2 \\
 & Product & 95.0 & 94.5 & 94.1 & 92.3 \\
 & Maximum & 94.2 & 93.7 & 94.1 & 92.0 \\
\hline
16-way-split & Summation & 93.7 & 95.5 & 92.6 & 91.26 \\
 & Product & 93.9 & 95.6 & 93.0 & 91.3 \\
 & Maximum & 93.7 & 95.5 & 93.7 & 90.4 \\
\hline
\end{tabular}
\caption{Image Level}
\end{table}

\begin{table}[h]
\centering  
\begin{tabular}{|c|c|c|c|c|c|}
\hline 
Segmentation & Fusion & \multicolumn{4}{|c|}{Magnification Factor} \\
\cline{3-6} 
& & 40 & 100 & 200 & 400 \\
\hline 
Non-split & - & 93.1 & 90.6 & 92.5 & 88.0 \\
\hline 
Quarter-split & Summation & 95.0 & 94.6 & 94.3 & 92.2 \\
 & Product & 94.9 & 94.2 & 94.5 & 92.3 \\
 & Maximum & 94.2 & 93.3 & 94.4 & 91.8 \\
\hline 
16-way-split & Summation & 94.0 & 94.1 & 91.4 & 91.2 \\
 & Product & 94.2 & 94.8 & 91.9 & 91.3 \\
 & Maximum & 94.0 & 94.6 & 92.8 & 91.0 \\
\hline
\end{tabular}
\caption{Patient Level}
\end{table}

To further evaluate the effectiveness of the combination of image segmentation methods with fusion algorithms, the accuracy of the classification is compared with other algorithm used in research[5,10-12]. Our approach improved image recognition accuracy by 9.9\% to 12.2\%. compared to literature\cite{spanhol2015dataset}, in which the features are extracted manually and by traditional machine learning algorithm. Compared to the method in literature \cite{zhang2024enhanced} which utilizes convolutional neural networks for feature extraction and employs data augmentation techniques such as rotation and flipping, our approach improved image recognition rates by 2.3\% to 5.1\%. Compared to the method in literature \cite{zhang2024enhanced} which utilizes convolutional neural networks for feature extraction and employs sliding window data augmentation, our approach improved image recognition rates by 5\% to 11.5\%. Compared to the method in literature that utilizes deep feature extraction and feature fusion, our approach improved image recognition rates by 8.2\% to 10.9\%. Compared to the method in literature that employs a multi-task convolutional neural network, our approach improved image recognition rates by 10.2\% to 12\%. Furthermore, through the comparison of experimental results, it can be concluded that methods based on convolutional neural network feature extraction outperform those relying on manual feature extraction in terms of classification performance. Manual feature extraction not only requires professional personnel but is also time-consuming. Feature extraction based on convolutional neural network greatly improves the efficiency, and the accuracy is correspondingly improved. The features of convolutional neural networks, such as hierarchical feature learning and parameter sharing mechanisms and data enhancement capabilities, can more effectively capture image features than manual feature extraction. This advantage may be the key to improving efficiency and accuracy. Naturally, the validation of volume height accuracy is constrained by the limited dataset; hence, a substantial increase in data volume is essential for further verification. 

\section{Conclusion}
With the advancement of deep learning technology in the field of computing, image recognition technology is applied in various fields, including assisting clinicians in classifying tumor cells. In this study, deep learning technology was applied to the cell classification of breast cancer. Drawing conclusions based on the aforementioned experimental data results and data comparisons, the following findings listed as below:
(1)	The image segmentation methods significantly enhance the recognition accuracy of breast images. The performance of 4-way split is higher than that of 16-way split.
(2)	Different fusion algorithms have a relatively minor impact on the recognition rate of segmented images. Through comparisons with other relevant research, the efficacy of feature extraction through deep learning techniques has been validated to be superior to visually based manual feature extraction
To further improve the algorithm, it is considered to optimize image feature extraction methods, such as employing multiple models for feature extraction combined with feature fusion. By optimizing image segmentation methods and data augmentation techniques, further improvements in classification accuracy can be achieved. Building upon binary classification, there will be efforts to further categorize benign and malignant tumors, achieving finer segmentation.

\bibliographystyle{ACM-Reference-Format}

\bibliography{references}


\begin{thebibliography}{19}


\ifx \showCODEN    \undefined \def \showCODEN     #1{\unskip}     \fi
\ifx \showDOI      \undefined \def \showDOI       #1{#1}\fi
\ifx \showISBNx    \undefined \def \showISBNx     #1{\unskip}     \fi
\ifx \showISBNxiii \undefined \def \showISBNxiii  #1{\unskip}     \fi
\ifx \showISSN     \undefined \def \showISSN      #1{\unskip}     \fi
\ifx \showLCCN     \undefined \def \showLCCN      #1{\unskip}     \fi
\ifx \shownote     \undefined \def \shownote      #1{#1}          \fi
\ifx \showarticletitle \undefined \def \showarticletitle #1{#1}   \fi
\ifx \showURL      \undefined \def \showURL       {\relax}        \fi
\providecommand\bibfield[2]{#2}
\providecommand\bibinfo[2]{#2}
\providecommand\natexlab[1]{#1}
\providecommand\showeprint[2][]{arXiv:#2}

\bibitem[Bayramoglu et~al\mbox{.}(2016)]%
        {bayramoglu2016deep}
\bibfield{author}{\bibinfo{person}{Neslihan Bayramoglu}, \bibinfo{person}{Juho Kannala}, {and} \bibinfo{person}{Janne Heikkil{\"a}}.} \bibinfo{year}{2016}\natexlab{}.
\newblock \showarticletitle{Deep learning for magnification independent breast cancer histopathology image classification}. In \bibinfo{booktitle}{\emph{2016 23rd International conference on pattern recognition (ICPR)}}. IEEE, \bibinfo{pages}{2440--2445}.
\newblock


\bibitem[Belsare et~al\mbox{.}(2015)]%
        {belsare2015classification}
\bibfield{author}{\bibinfo{person}{AD Belsare}, \bibinfo{person}{MM Mushrif}, \bibinfo{person}{MA Pangarkar}, {and} \bibinfo{person}{N Meshram}.} \bibinfo{year}{2015}\natexlab{}.
\newblock \showarticletitle{Classification of breast cancer histopathology images using texture feature analysis}. In \bibinfo{booktitle}{\emph{Tencon 2015-2015 IEEE Region 10 Conference}}. IEEE, \bibinfo{pages}{1--5}.
\newblock


\bibitem[Chhikara and Parang(2023)]%
        {chhikara2023global}
\bibfield{author}{\bibinfo{person}{Bhupender~S Chhikara} {and} \bibinfo{person}{Keykavous Parang}.} \bibinfo{year}{2023}\natexlab{}.
\newblock \showarticletitle{Global Cancer Statistics 2022: the trends projection analysis}.
\newblock \bibinfo{journal}{\emph{Chemical Biology Letters}} \bibinfo{volume}{10}, \bibinfo{number}{1} (\bibinfo{year}{2023}), \bibinfo{pages}{451--451}.
\newblock


\bibitem[Cserni(2020)]%
        {cserni2020histological}
\bibfield{author}{\bibinfo{person}{G{\'a}bor Cserni}.} \bibinfo{year}{2020}\natexlab{}.
\newblock \showarticletitle{Histological type and typing of breast carcinomas and the WHO classification changes over time}.
\newblock \bibinfo{journal}{\emph{Pathologica}} \bibinfo{volume}{112}, \bibinfo{number}{1} (\bibinfo{year}{2020}), \bibinfo{pages}{25}.
\newblock


\bibitem[Elmore et~al\mbox{.}(2015)]%
        {elmore2015diagnostic}
\bibfield{author}{\bibinfo{person}{Joann~G Elmore}, \bibinfo{person}{Gary~M Longton}, \bibinfo{person}{Patricia~A Carney}, \bibinfo{person}{Berta~M Geller}, \bibinfo{person}{Tracy Onega}, \bibinfo{person}{Anna~NA Tosteson}, \bibinfo{person}{Heidi~D Nelson}, \bibinfo{person}{Margaret~S Pepe}, \bibinfo{person}{Kimberly~H Allison}, \bibinfo{person}{Stuart~J Schnitt}, {et~al\mbox{.}}} \bibinfo{year}{2015}\natexlab{}.
\newblock \showarticletitle{Diagnostic concordance among pathologists interpreting breast biopsy specimens}.
\newblock \bibinfo{journal}{\emph{Jama}} \bibinfo{volume}{313}, \bibinfo{number}{11} (\bibinfo{year}{2015}), \bibinfo{pages}{1122--1132}.
\newblock


\bibitem[Guo and Wen(2021)]%
        {guo2021multi}
\bibfield{author}{\bibinfo{person}{Qianyu Guo} {and} \bibinfo{person}{Jing Wen}.} \bibinfo{year}{2021}\natexlab{}.
\newblock \showarticletitle{Multi-level Fusion Based Deep Convolutional Network for Image Quality Assessment}. In \bibinfo{booktitle}{\emph{Pattern Recognition. ICPR International Workshops and Challenges: Virtual Event, January 10--15, 2021, Proceedings, Part VI}}. Springer, \bibinfo{pages}{670--678}.
\newblock


\bibitem[Hu et~al\mbox{.}(2022)]%
        {hu2022design}
\bibfield{author}{\bibinfo{person}{Zhirui Hu}, \bibinfo{person}{Jinyang Li}, \bibinfo{person}{Zhenyu Pan}, \bibinfo{person}{Shanglin Zhou}, \bibinfo{person}{Lei Yang}, \bibinfo{person}{Caiwen Ding}, \bibinfo{person}{Omer Khan}, \bibinfo{person}{Tong Geng}, {and} \bibinfo{person}{Weiwen Jiang}.} \bibinfo{year}{2022}\natexlab{}.
\newblock \showarticletitle{On the design of quantum graph convolutional neural network in the nisq-era and beyond}. In \bibinfo{booktitle}{\emph{2022 IEEE 40th International Conference on Computer Design (ICCD)}}. IEEE, \bibinfo{pages}{290--297}.
\newblock


\bibitem[Konovalenko et~al\mbox{.}(2022)]%
        {konovalenko2022influence}
\bibfield{author}{\bibinfo{person}{Ihor Konovalenko}, \bibinfo{person}{Pavlo Maruschak}, \bibinfo{person}{Halyna Kozbur}, \bibinfo{person}{Janette Brezinov{\'a}}, \bibinfo{person}{Jakub Brezina}, \bibinfo{person}{Bohdan Nazarevich}, {and} \bibinfo{person}{Yaroslav Shkira}.} \bibinfo{year}{2022}\natexlab{}.
\newblock \showarticletitle{Influence of uneven lighting on quantitative indicators of surface defects}.
\newblock \bibinfo{journal}{\emph{Machines}} \bibinfo{volume}{10}, \bibinfo{number}{3} (\bibinfo{year}{2022}), \bibinfo{pages}{194}.
\newblock


\bibitem[Li et~al\mbox{.}(2018)]%
        {li2018graph}
\bibfield{author}{\bibinfo{person}{Ruoyu Li}, \bibinfo{person}{Jiawen Yao}, \bibinfo{person}{Xinliang Zhu}, \bibinfo{person}{Yeqing Li}, {and} \bibinfo{person}{Junzhou Huang}.} \bibinfo{year}{2018}\natexlab{}.
\newblock \showarticletitle{Graph CNN for survival analysis on whole slide pathological images}. In \bibinfo{booktitle}{\emph{International Conference on Medical Image Computing and Computer-Assisted Intervention}}. Springer, \bibinfo{pages}{174--182}.
\newblock


\bibitem[Li et~al\mbox{.}(2024)]%
        {10440308}
\bibfield{author}{\bibinfo{person}{Sai Li}, \bibinfo{person}{Peng Kou}, \bibinfo{person}{Miao Ma}, \bibinfo{person}{Haoyu Yang}, \bibinfo{person}{Shuo Huang}, {and} \bibinfo{person}{Zhengyi Yang}.} \bibinfo{year}{2024}\natexlab{}.
\newblock \showarticletitle{Application of Semi-Supervised Learning in Image Classification: Research on Fusion of Labeled and Unlabeled Data}.
\newblock \bibinfo{journal}{\emph{IEEE Access}}  \bibinfo{volume}{12} (\bibinfo{year}{2024}), \bibinfo{pages}{27331--27343}.
\newblock
\urldef\tempurl%
\url{https://doi.org/10.1109/ACCESS.2024.3367772}
\showDOI{\tempurl}


\bibitem[Liu et~al\mbox{.}(2023a)]%
        {liu2023unveiling}
\bibfield{author}{\bibinfo{person}{Yongfei Liu}, \bibinfo{person}{Haoyu Yang}, {and} \bibinfo{person}{Chenwei Wu}.} \bibinfo{year}{2023}\natexlab{a}.
\newblock \showarticletitle{Unveiling patterns: A study on semi-supervised classification of strip surface defects}.
\newblock \bibinfo{journal}{\emph{IEEE Access}}  \bibinfo{volume}{11} (\bibinfo{year}{2023}), \bibinfo{pages}{119933--119946}.
\newblock


\bibitem[Liu et~al\mbox{.}(2023b)]%
        {liu2023ising}
\bibfield{author}{\bibinfo{person}{Zhuo Liu}, \bibinfo{person}{Yunan Yang}, \bibinfo{person}{Zhenyu Pan}, \bibinfo{person}{Anshujit Sharma}, \bibinfo{person}{Amit Hasan}, \bibinfo{person}{Caiwen Ding}, \bibinfo{person}{Ang Li}, \bibinfo{person}{Michael Huang}, {and} \bibinfo{person}{Tong Geng}.} \bibinfo{year}{2023}\natexlab{b}.
\newblock \showarticletitle{Ising-cf: A pathbreaking collaborative filtering method through efficient ising machine learning}. In \bibinfo{booktitle}{\emph{2023 60th ACM/IEEE Design Automation Conference (DAC)}}. IEEE, \bibinfo{pages}{1--6}.
\newblock


\bibitem[Lu et~al\mbox{.}(2018)]%
        {lu2018nuclear}
\bibfield{author}{\bibinfo{person}{Cheng Lu}, \bibinfo{person}{David Romo-Bucheli}, \bibinfo{person}{Xiangxue Wang}, \bibinfo{person}{Andrew Janowczyk}, \bibinfo{person}{Shridar Ganesan}, \bibinfo{person}{Hannah Gilmore}, \bibinfo{person}{David Rimm}, {and} \bibinfo{person}{Anant Madabhushi}.} \bibinfo{year}{2018}\natexlab{}.
\newblock \showarticletitle{Nuclear shape and orientation features from H\&E images predict survival in early-stage estrogen receptor-positive breast cancers}.
\newblock \bibinfo{journal}{\emph{Laboratory investigation}} \bibinfo{volume}{98}, \bibinfo{number}{11} (\bibinfo{year}{2018}), \bibinfo{pages}{1438--1448}.
\newblock


\bibitem[Lu et~al\mbox{.}(2023)]%
        {lu2023transflow}
\bibfield{author}{\bibinfo{person}{Yawen Lu}, \bibinfo{person}{Qifan Wang}, \bibinfo{person}{Siqi Ma}, \bibinfo{person}{Tong Geng}, \bibinfo{person}{Yingjie~Victor Chen}, \bibinfo{person}{Huaijin Chen}, {and} \bibinfo{person}{Dongfang Liu}.} \bibinfo{year}{2023}\natexlab{}.
\newblock \showarticletitle{Transflow: Transformer as flow learner}. In \bibinfo{booktitle}{\emph{Proceedings of the IEEE/CVF Conference on Computer Vision and Pattern Recognition}}. \bibinfo{pages}{18063--18073}.
\newblock


\bibitem[Ma et~al\mbox{.}(2023)]%
        {ma2023implementation}
\bibfield{author}{\bibinfo{person}{Danqing Ma}, \bibinfo{person}{Bo Dang}, \bibinfo{person}{Shaojie Li}, \bibinfo{person}{Hengyi Zang}, {and} \bibinfo{person}{Xinqi Dong}.} \bibinfo{year}{2023}\natexlab{}.
\newblock \showarticletitle{Implementation of computer vision technology based on artificial intelligence for medical image analysis}.
\newblock \bibinfo{journal}{\emph{International Journal of Computer Science and Information Technology}} \bibinfo{volume}{1}, \bibinfo{number}{1} (\bibinfo{year}{2023}), \bibinfo{pages}{69--76}.
\newblock


\bibitem[Spanhol et~al\mbox{.}(2017)]%
        {spanhol2017deep}
\bibfield{author}{\bibinfo{person}{Fabio~A Spanhol}, \bibinfo{person}{Luiz~S Oliveira}, \bibinfo{person}{Paulo~R Cavalin}, \bibinfo{person}{Caroline Petitjean}, {and} \bibinfo{person}{Laurent Heutte}.} \bibinfo{year}{2017}\natexlab{}.
\newblock \showarticletitle{Deep features for breast cancer histopathological image classification}. In \bibinfo{booktitle}{\emph{2017 IEEE International Conference on Systems, Man, and Cybernetics (SMC)}}. IEEE, \bibinfo{pages}{1868--1873}.
\newblock


\bibitem[Spanhol et~al\mbox{.}(2015)]%
        {spanhol2015dataset}
\bibfield{author}{\bibinfo{person}{Fabio~A Spanhol}, \bibinfo{person}{Luiz~S Oliveira}, \bibinfo{person}{Caroline Petitjean}, {and} \bibinfo{person}{Laurent Heutte}.} \bibinfo{year}{2015}\natexlab{}.
\newblock \showarticletitle{A dataset for breast cancer histopathological image classification}.
\newblock \bibinfo{journal}{\emph{Ieee transactions on biomedical engineering}} \bibinfo{volume}{63}, \bibinfo{number}{7} (\bibinfo{year}{2015}), \bibinfo{pages}{1455--1462}.
\newblock


\bibitem[Spanhol et~al\mbox{.}(2016)]%
        {spanhol2016breast}
\bibfield{author}{\bibinfo{person}{Fabio~Alexandre Spanhol}, \bibinfo{person}{Luiz~S Oliveira}, \bibinfo{person}{Caroline Petitjean}, {and} \bibinfo{person}{Laurent Heutte}.} \bibinfo{year}{2016}\natexlab{}.
\newblock \showarticletitle{Breast cancer histopathological image classification using convolutional neural networks}. In \bibinfo{booktitle}{\emph{2016 international joint conference on neural networks (IJCNN)}}. IEEE, \bibinfo{pages}{2560--2567}.
\newblock


\bibitem[Zhang et~al\mbox{.}(2024)]%
        {zhang2024enhanced}
\bibfield{author}{\bibinfo{person}{Fei Zhang}, \bibinfo{person}{Mingxuan Xiao}, \bibinfo{person}{Weimin Wang}, \bibinfo{person}{Yufeng Li}, {and} \bibinfo{person}{Xu Yan}.} \bibinfo{year}{2024}\natexlab{}.
\newblock \showarticletitle{Enhanced Breast Cancer Classification through Data Fusion Modeling}.
\newblock \bibinfo{journal}{\emph{Journal of Theory and Practice of Engineering Science}} \bibinfo{volume}{4}, \bibinfo{number}{01} (\bibinfo{year}{2024}), \bibinfo{pages}{79--85}.
\newblock


\end{thebibliography}

\end{document}